\documentstyle[11pt]{article}

\begin{document}
\oddsidemargin .1in
\evensidemargin 0 true pt
\topmargin -.4in


\def\ra{{\rightarrow}}
\def\a{{\alpha}}
\def\b{{\beta}}
\def\l{{\lambda}}
\def\eps{{\epsilon}}
\def\T{{\Theta}}
\def\t{{\theta}}
\def\co{{\cal O}}
\def\car{{\cal R}}
\def\caf{{\cal F}}
\def\cs{{\Theta_S}}
\def\pr{{\partial}}
\def\tri{{\triangle}}
\def\na{{\nabla }}
\def\S{{\Sigma}}
\def\s{{\sigma}}
\def\sp{\vspace{.15in}}
\def\hs{\hspace{.25in}}

\newcommand{\be}{\begin{equation}} \newcommand{\ee}{\end{equation}}
\newcommand{\bea}{\begin{eqnarray}}\newcommand{\eea}
{\end{eqnarray}}


\begin{titlepage}
\topmargin= -.2in
\textheight 9.5in

\begin{center}
\baselineskip= 17 truept

\vspace{.3in}
\centerline{\Large\bf Gravity dual $D_3$-braneworld and Open/Closed string duality}


\vspace{.6in}

\noindent
{{{\large Supriya Kar}\footnote{skkar@physics.du.ac.in }},
{{\large K. Priyabrat Pandey}\footnote{kppandey@physics.du.ac.in }},
{{\large Abhishek K. Singh}\footnote{abhishek@physics.du.ac.in }} {{\large and Sunita Singh}\footnote{sunita@physics.du.ac.in }}}

\vspace{.2in}

\noindent
{\large Department of Physics \& Astrophysics\\
{\large University of Delhi, New Delhi 110 007, India}}

\vspace{.2in}

{\today}
\thispagestyle{empty}

\vspace{.6in}
\begin{abstract}

\vspace{.2in}
A covariantly constant dynamical two-form is exploited on a $D_3$-brane to obtain its gravity dual action, governing an $S^3$ deformed $AdS_5$ black hole, in a type IIB string theory on $S^1\times K3$. We invoke the Kaluza-Klein compactification to work out the 
open/closed string duality. Interestingly, the Reissner-Nordstrom black hole is obtained on the ``non-Riemannian'' braneworld.

\end{abstract}
\end{center}

\vspace{.2in}

\baselineskip= 14 truept

\vspace{1in}



\end{titlepage}

\baselineskip= 18 truept

\section{Introduction}
In General Theory of Relativity (GTR), the space-time curvatures are based on the intrinsic notion of geometry, by incorporating a nontrivial metric tensor on a (Riemannian) manifold. The fact that the space-time curvatures are governed by the nonlinear matter field, enforces hidden gravity/gauge duality in Einstein's formulation. The duality becomes evident in presence of a fifth dimension, $i.e.$ in Kaluza-Klein gravity theory. Importantly, the duality takes a precise form in $AdS$ gravity and is established as a familiar bulk/boundary or AdS/CFT correspondence \cite{maldacena} in string theory.

\vspace{.15in}
In the context, non-perturbative Dirichlet (D-) branes \cite{polchinski} are believed to enlighten the aspects of geometry/matter duality coded
in Einstein's formulation. In fact, the nonlinear $U(1)$ gauge dynamics established on a $D$-brane \cite{seiberg-witten} is indeed remarkable. Along with the higher derivative corrections, the nonlinear gauge sector is believed to govern the space-time curvature. Most importantly, the underlying duality has been considerably explored through the success of $5D$ Anti de Sitter ($AdS_5$) gravity and $D_3$-brane gauge theory correspondence established in string theory. Interestingly, the non-commutative open string (NCOS) \cite{minwalla} leading to a deformed $D$-braneworld, its gravity correspondence \cite{li} and various (non-linear) charged $D_3$-braneworld black holes \cite{mars}-\cite{kar} 
are established in the recent years.

\vspace{.15in}
In the paper, we attempt to formalize the nonlinear $U(1)$ gauge dynamics on a $D_3$-brane in terms of its geometric dual in a string theory. We explore a $D_3$-braneworld in a plausible fundamental open string formulation and obtain a closed string dual of the $D_3$-braneworld, $i.e.$ a curved $D_3$-braneworld in $5D$. We consider a covariantly constant dynamical two-form to incorporate new connections, $i.e.$ a torsion, which in turn lead to an unique modified covariant derivative on the curved $D_3$-braneworld. Our starting point is a flat $D_3$-brane, which is known to govern the gravity decoupling limit $E\rightarrow E_c= {{g}\over{2\pi\alpha'}}$ in an open string theory. We revisit the decoupling limit as a cancellation of torsion arising from the zeroth order term in $B$-field to that of all higher orders, keeping the geometrical notion of Einstein's formulation. In fact, an asymptotic $D_3$-brane may be thought of as a transition world-volume connecting the underlying NCOS and a fundamental open string theory. As a result, the $D_3$-braneworld in the classical regime, $i.e.\ g\rightarrow 0$ and $\alpha'\rightarrow 0$, may be seen to be described by the nontrivial generic curvatures underlying a (modified) open string theory.

\section{Two-form as a source of gravity}

\subsection{New Connections}
The covariantly constant dynamical $B_{\mu\nu}$ source implies that non-trivial connections ${\mathbf\Gamma}_{\mu\nu\lambda}$ do exist on the $D_3$-braneworld. Since Einstein's gravity decouples, the generic curvature on a curved $D_3$-braneworld may be seen to be governed by the ``non-Riemannian'' geometries. The vanishing field strength for the dynamical matter introduces the notion of new geometries on the braneworld. 
underlying a modified open string theory. Formally, the field strength may be expressed as
\be
{\cal H}_{\mu\nu\lambda}=\left ({\cal D}_{\mu}B_{\nu\lambda} + {\bar F}_{\mu\nu}A_{\lambda}\right ) + {\rm cyclic}\ ({\rm in}\ \mu,\nu,\lambda) \ ,\label{field-1}
\ee
where the $U(1)$ field strength ${\bar F}_{\mu\nu}= (2\pi\alpha')F_{\mu\nu}$. The covariant derivative ${\cal D}_{\mu}$ takes into account a non-dynamical metric $g_{\mu\nu}$ and a dynamical two-form $B_{\mu\nu}$ and ${\cal D}_{\lambda}g_{\mu\nu} = 0$.
We rewrite the new connection as
\be
{\mathbf\Gamma}_{\mu\nu\lambda}= {\mathbf\Gamma}^{(0)}_{\mu\nu\lambda}-{\mathbf\Gamma}^{(h)}_{\mu\nu\lambda} + \Big ({\bar F}_{\mu\nu}A_{\lambda}+
{\rm cyclic}\Big )\ ,\label{connect3}
\ee
where ${\mathbf\Gamma}^{(0)}_{\mu\nu\lambda}=-{1\over2}\left ( \partial_{\mu}B_{\nu\lambda} + {\rm cyclic}\right )$ and
${\mathbf\Gamma}^{(h)}_{\mu\nu\lambda}=- ( {\mathbf\Gamma}^{\rho}_{\mu\nu}B_{\rho\lambda} + {\rm cyclic} )$, respectively, 
denote the zeroth and higher orders, in two-form, defined on a $D_3$-braneworld. The gravity decoupling limit, in the formalism, is defined  
by ${\mathbf\Gamma}^{(0)}_{\mu\nu\lambda}\rightarrow {\mathbf\Gamma}^{(h)}_{\mu\nu\lambda}$. In the low energy regime, the non-Riemannian 
geometry on a $4D$ braneworld may be viewed in terms of two equivalent open string descriptions. They are: (i) a matter dominated braneworld for $|{\mathbf\Gamma}^{(0)}_{\mu\nu\lambda}|>>|{\mathbf\Gamma}^{(h)}_{\mu\nu\lambda}|$ and (ii) a gravity or geometric braneworld for
$|{\mathbf\Gamma}^{(h)}_{\mu\nu\lambda}|>>|{\mathbf\Gamma}^{(0)}_{\mu\nu\lambda}|$. The matter dominated braneworld precisely corresponds to
the non-commutative open string (NCOS) description \cite{minwalla}. On the other hand, the geometric braneworld may be seen to govern by a (modified) fundamental open string. The underlying open string, in an Einstein gravity decoupled braneworld, is a new phenomenon in the formalism.

\subsection{Gravity and gauge curvatures}
The generic gravity and gauge curvatures on the $D_3$-braneworld may be computed from the commutator of appropriate gauge covariant derivatives on it. In presence of a gauge field, the appropriate covariant derivative becomes $D_{\mu}= \left ({\cal D}_{\mu} - A_{\mu}\right )$. A priori, the nontrivial curvatures on a $D_3$-braneworld with may be given by
\be
\Big [ D_{\mu} , D_{\nu} \Big ]A_{\lambda}= {\cal F}_{\nu\mu}A_{\lambda}+ {K_{\mu\nu\lambda}}^{\rho}A_{\rho} + T^{\rho}_{\mu\nu}\ {\cal D}_{\rho}A_{\lambda} + {1\over2}\left ( \Gamma^{\rho}_{\mu\sigma} T^{\sigma}_{\nu\lambda} +
\Gamma^{\sigma}_{\nu\lambda} T^{\rho}_{\mu\sigma} -\Gamma^{\sigma}_{\mu\lambda} T^{\rho}_{\nu\sigma} -\Gamma^{\rho}_{\nu\sigma} T^{\sigma}_{\mu\lambda}\right )A_{\rho}\ .\label{curvature-1}
\ee
The non-linear $U(1)$ field strength ${\cal F}_{\mu\nu}=F_{\mu\nu} + T^{\rho}_{\mu\nu}A_{\rho}$,
where $T^{\rho}_{\mu\nu}=-2{\mathbf\Gamma}^{\sigma}_{\mu\nu}$ is a torsion on the braneworld. 
The non-Riemannian curvature tensor is precisely due to the two-form and takes an identical form to the Riemannian-Christoffel curvature 
tensor $R_{\mu\nu\lambda\rho}$. It is given by
\be
{K_{\mu\nu\lambda}}^{\rho}=\partial_{\nu} {\mathbf\Gamma}^{\rho}_{\mu\lambda} - \partial_{\mu}{\mathbf\Gamma}^{\rho}_{\nu\lambda} + {\mathbf\Gamma}^{\sigma}_{\mu\lambda}
{\mathbf\Gamma}^{\rho}_{\nu\sigma}-{\mathbf\Gamma}^{\sigma}_{\nu\lambda}{\mathbf\Gamma}^{\rho}_{\mu\sigma}\ .\label{tensor}
\ee
Importantly for a constant torsion, $K_{\mu\nu\lambda\rho}\rightarrow R_{\mu\nu\lambda\rho}$. Incorporating the irreducible generic non-Riemannian curvature tensor, we rewrite, the $D_3$-braneworld action as
\be
S= {1\over{a}}\int\ d^4x {\sqrt{-g}}\ \left ( K - {a\over4}\ F^2\right )\ .\label{D3-action}
\ee
A priori, simple looking braneworld may be seen to be governed by the intrinsically coupled gravity and gauge curvatures from the
$B_{\mu\nu}$ and $A_{\mu}$ equations of motion. They are, respectively, given by
\bea
&&\partial_{\lambda}\Gamma^{\lambda\mu\nu} - {1\over2} \Big ( g^{\alpha\beta}\partial_{\lambda}\ g_{\alpha\beta}\Big ) \Gamma^{\lambda\mu\nu}=0
\nonumber\\
{\rm and}&& a\left (\nabla_{\lambda}F^{\lambda\mu}\right ) - 6 F_{\lambda\nu}{\mathbf\Gamma}^{\lambda\nu\mu} + 12 \nabla_{\lambda}
\left ({\mathbf\Gamma}^{\lambda\mu\nu}A_{\nu}\right )=0 .\label{5d-eq}
\eea
\subsection{Reissner-Nordstrom vacua on braneworld}
In this section, we address a non-Riemannian formulation leading to Reissner-Nordstrom black hole geometry on the $D_3$-braneworld. 
The anstaz for a two-form and a gauge field leading to a charged black hole may be given by
$B_{0\phi}= - M\cos \theta$ and $A_0= Q_e/r$,
where $M$ and $Q_e$ are constants and shall be identified with the mass and the electric charge. 
The gauge choice leads to a nontrivial connection ${\mathbf\Gamma}_{0\theta\phi}= {M\over2} \sin\theta$ and an extremal charged black hole geometry may be seen to emerge through an effective metric $G_{\mu\nu}= g_{\mu\nu}-B_{\mu\lambda}g^{\lambda\rho}g_{\rho\nu}$ description in an open string theory. However, we consider a (generalized) geodesic expansion for the effective metric components $G_{00}=G^{rr}$, around an $S^2$-symmetric vacua, in a non-relativistic limit. 
The $G_{00}$ component of the effective metric may be obtained from a geodesic expansion represented by the derivative corrections of gauge potentials of increasing rank in each term. It is given by
$G_{00}= (g_{00} + h^{(0)}_{00} + h^{(1)}_{00} + h^{(2)}_{00} + \dots )$. The nontrivial scalar quantities in 2nd, 3rd and 4th terms are essentially the distinct contributions, respectively, out of the scalar, vector and two-form fields. To avoid an ambiguous form of $h^{(2)}_{00}$, we rather consider a Poincare dual of $B$-field, $i.e.$ an axion field $\chi(x)$ as a source of Newtonian potential on the braneworld.
The axion and gauge gravitational potentials are, respectively, worked out to yield $h^{(0)}_{00}= -2M/r$ and $h^{(1)}_{00}= Q_e^2/r^2$.
The axion source on the $D_3$-braneworld leads to an electric charged black hole which in turn may be viewed in terms of both the charges
($Q^2=Q_e^2+Q_m^2$) in a boosted coordinate system. In the case, the ``non-Riemannian'' $D_3$-braneworld may be 
seen to describe a typical Reissner-Nordstrom black hole geometry known in Riemannian geometry. The nontrivial vacua common to two distinct tensor formulations of gravity in string theory is remarkable. It may provide clues to enhance our understanding of de Sitter vacua.
\section{Gravity dual $D_3$-braneworld}

The gravity and gauge curvatures on the curved $D_3$-braneworld in $4D$ may be seen to have its origin in a five dimensional gravity theory governed by a dynamical $B$-field in a truncated closed string theory. A priori, one may begin with the type IIB string theory on $S^1\times K3$. 
We consider a non-dynamical metric and for simplicity, focus on constant scalars and gauge fields. The generic the notion of closed string is maintained in $5D$ by the (covariantly constant) dynamical $B$-field in absence of the dynamical metric on the curved braneworld. The relevant closed string action may be given by
\be
S= \int d^5y\ {\sqrt{-{\tilde g}}}\ {\tilde K}\ ,\label{5d-action}
\ee
where ${\tilde g}_{\mu\nu}$ is a constant metric in the Cartesian coordinates. An ansatz for the $B$-field, around a $S^3$-symmetric vacua, is worked out to yield
\be
B_{0\psi}= B_{r\psi}=r_0\ ,\qquad B_{\theta\psi}= q (\cot\theta\sin^2\psi)\qquad {\rm and}\quad B_{\psi\phi}= q
(\cos\theta\sin^2\psi)\ ,\label{5d-ansatz}
\ee
where $0<\theta\le\pi$, $0<\psi\le\pi$, $0\le\phi\le 2\pi$ and ($r_0$ and $q$) are are arbitrary constants. The nontrivial connections 
are given by
\be
{\mathbf\Gamma}_{\theta\psi\phi}= {{q}\over{2}}(\sin\theta\sin^2\psi)\ {\rm and}\quad 
{\mathbf\Gamma}_{\theta 0\phi}= {\mathbf\Gamma}_{\theta r\phi} = {{(qr_0)}\over{2r^2}}(\sin\theta\sin^2\psi)\ .\label{5d-ansatz-2}
\ee
Importantly, the $D$-braneworld tension turns out to be a significant vacuum energy density in the effective metric description. As a result, the radial coordinate picks up an asymptotic 
cut-off $b$, the AdS radius, in the curved $D$-braneworld formalism. The $5D$ black hole metric may be constructed using the $B$-field ansatz around an AdS vacua. The $AdS_5$ geometry may be seen to be described by an $AdS_2$, and a, $S^2$, deformed $S^3$-symmetry in the closed string theory.
It is given by
\be
ds^2=-\left (1+{{r^2}\over{b^2}}-{{r_0^2}\over{r^2}}\right ) dt^2 + \left ( 1 + {{r^2}\over{b^2}} -{{r_0^2}\over{r^2}}\right )^{-1} dr^2
+r^2 \left ( 1+ {{2f^2}\over{r^4}}\right ) d\Omega_3^2 - {{B^2_{\theta\psi}}\over{r^2}} d\Omega_2^2\ ,\label{5d-ads}
\ee
where $f= q(\cot\theta\sin\psi)$.
A priori, the fundamental closed string action (\ref{5d-action}), may alternately be viewed in an effective metric description to yield the usual Kaluza-Klein gravity in presence of a cosmological constant. In that case, the geometry may be seen to be governed by the Schwarzschild $AdS_5$ black hole with $S_3$-symmetry \cite{sen,horowitz-polchinski}. In other words, $AdS_5$ black hole in the $B$-field theory of gravity governs an deformed $S^3$, where as the $3D$-spherical symmetry is preserved in the metric source theory of gravity. As a result, the $D_3$-braneworld action (\ref{5d-action}) possibly governs a generalized gravity dynamics, which is beyond a typical $5D$ Kaluza-Klein gravity theory.
For a constant $\Gamma_{\theta\psi\phi}\neq 0$, the conformal factor and the $S^3$ deformation geometry become trivial in the $AdS_5$ black hole (\ref{5d-ads}). In the case, the Reimannian notion of manifold becomes manifest and the geometry is in agreement with the Schwrazschild $AdS_2\times S^3$ black hole obtained in the Kaluza-Klein theory.

\vspace{.15in}
The Kaluza-Klein compactification of the fifth dimension in the closed string theory (\ref{5d-action}), in a static gauge, precisely reduces to the gravity dual $D_3$-brane action (\ref{D3-action}). For instance, if $\psi$ denotes a compactified coordinate, the gauge potential may be worked out in $4D$, from the $B$-field ansatz, to yield 
\be
A_{\mu}=\Big (r_0,\ r_0,\ q \cot\theta,\ -q\cos\theta\Big )\ .\label{4d-gauge}
\ee
The linear electromagnetic field, in the gravity dual $D_3$-braneworld, is a priori governed by the radial component of the magnetic field $B_r= q/r^2$, along with a nontrivial $B$-field in a Kaluza-Klein compactified closed string theory{\footnote{The Kaluza-Klein compactification may  also be seen to yield an irreducible curvature precisely due to a generic nonlinear $U(1)$ gauge field in the theory.}}. It is thought provoking to understand the Kaluza-Klein compactification in a $B$-field theory of gravity as a $T$-duality between a, fundamental, open/closed string. It may also be understood as a (bulk) gravity/(boundary) gauge theory correspondence. 
In fact, the $D_3$-braneworld action possesses a dual notion in a type IIB string theory. They are: (i) a fundamental closed string description in $5D$ and (ii) a fundamental open string in $4D$. The geometric braneworld curvature in $4D$ essentially confirms the fundamental nature of open string there. 

\vspace{.15in}
Interestingly, all the higher derivative corrections vanish for a self-dual electro-magnetic field \cite{gibbons}.
As a result, the deformed $AdS_5$ (\ref{5d-ads}) in a closed string theory may be seen to reduce to an $AdS_4$ black hole whose curvature is due to the nonlinear electric/magnetic charge ($q$) in an open string theory obtained by one of the author in ref.\cite{kar}.

\section{Concluding remarks}
The gravity dual of a $D_3$-brane action is obtained in a type IIB string theory on $S^1\times K3$. The geometrical description on the braneworld confirms a dual nature of two-form, $i.e.$ a closed string description in $5D$ and a NCOS description in $4D$. The fact is reassured by its underlying deformed $AdS$ black holes geometries both in closed and open string theories.
As a bonus, we obtain a fundamental open string theory $T$-dual to a closed string theory in $5D$. Our analysis leads to the birth of a generic curvature tensor $K_{\mu\nu\lambda\rho}$ which retains the properties of Riemann-Christoffel tensor except for the symmetric property under the exchange of its pair of indices. However for a constant torsion, the gravity dual braneworld may be seen to be described by the 
Riemannian geometry.

\vspace{.15in}
A constant two-form ansatz may be obtained for $q=0$ in the gravity dual $D_3$-braneworld at the expense of an effective dynamical theory of gravity. It restores the $S^3$-symmetry and describes an $AdS_5$ Schwarzschild black hole in a typical Kaluza-Klein gravity in presence of a cosmlogical constant. On the other hand, the global mode of a $B$-field in an open string theory is known to yield the deformation geometry \cite{seiberg-witten}. In fact, it may be interesting to investigate the thermal aspects associated with the non-commutative geometry in 
the curved $D$-braneworld along the spirit of entropic force of gravity discussed recently \cite{verlinde-10}. It confirms that the deformation geometry, in string theory, is intrinsic to the $B$-field source of gravity, which is unlike to a metric source of gravity. In other words, the open/closed string duality in presence of $B$-field retains the notion of the deformation geometry. Interestingly, the torsion does not break the $S^3$-symmetry in the $AdS_5$ black hole (\ref{5d-ads}), rather it is broken solely by the $B$-field.

\sp

\sp

\noindent
{\large\bf Acknowledgments}

\sp
A preliminary version of the research work was presented by S.K. in the Alok Kumar memorial meeting at IOP, Bhubaneswar during 17-19 January 2010.
The work of S.K. is partly supported by a research project under D.S.T, Govt.of India. AKS acknowledges C.S.I.R, New Delhi for a partial support.

\sp

\sp

\def\anp{Ann. of Phys.}
\def\prl{Phys.Rev.Lett.}
\def\prd#1{{Phys.Rev.}{\bf D#1}}
\def\jhep{JHEP}
\def\cqg#1{{Class.\& Quant.Grav.}}
\def\plb#1{{Phys. Lett.} {\bf B#1}}
\def\npb#1{{Nucl. Phys.} {\bf B#1}}
\def\mpl#1{{Mod. Phys. Lett} {\bf A#1}}
\def\ijmpa#1{{Int.J.Mod.Phys.}{\bf A#1}}
\def\rmp#1{{Rev. Mod. Phys.} {\bf 68#1}}


\end{document}